\documentclass[12pt]{article}

\usepackage{amsmath,amsfonts,amssymb}
\usepackage{graphicx}
\usepackage{subfigure}

\newcommand{\be}{\begin{equation}}
\newcommand{\ee}{\end{equation}}
\newcommand{\ba}{\begin{eqnarray}}
\newcommand{\ea}{\end{eqnarray}}

\begin{document}
\newcommand{\todo}[1]{{\em \small {#1}}\marginpar{$\Longleftarrow$}}
\newcommand{\labell}[1]{\label{#1}\qquad_{#1}} 

\rightline{DCPT-12/47}
\rightline{NSF-KITP-12-234}

\vskip 1cm

\begin{center} {\Large \bf Boundary conditions for scalars in Lifshitz}
\end{center}
\vskip 1cm

\renewcommand{\thefootnote}{\fnsymbol{footnote}}
\centerline{Tom\'as Andrade$^{a,b}$\footnote{tomas.andrade@durham.ac.uk} and Simon F. Ross$^b$\footnote{s.f.ross@durham.ac.uk} }
\vskip .5cm
\centerline{\it $^a$ Department of Physics, UCSB}
\centerline{\it Santa Barbara, CA, 93106, USA}
\vskip .5cm
\centerline{\it $^b$ Centre for Particle Theory, Department of Mathematical Sciences}
\centerline{\it Durham University, South Road, Durham DH1 3LE, U.K.}

\setcounter{footnote}{0}
\renewcommand{\thefootnote}{\arabic{footnote}}

\begin{abstract}
We consider the quantisation of scalar fields on a Lifshitz background, exploring the possibility of alternative boundary conditions, allowing the slow falloff mode to fluctuate. We show that the scalar field with alternative boundary conditions is normalizable for a larger range of masses than in the AdS case. However, we then find a new instability for alternative boundary conditions, implying that the range of masses where alternative boundary conditions define a well-behaved dual theory is $m_{BF}^2 < m^2 < m^2_{BF} +1$, analogously to the AdS case. The instability is of a novel type, with modes of arbitrarily large momentum which grow exponentially in time; it is therefore essentially a UV effect, and implies that the dual field theory is simply not defined where it appears. We discuss the interpretation in the dual field theory, and give a proposed lower bound on the dimension of scalar operators.
\end{abstract}

\section{Introduction}

The holographic description of field theories with anisotropic scaling symmetry presents an interesting extension of AdS/CFT, which may have valuable applications in condensed matter theory. The simplest example of such a dual description is the Lifshitz metric originally constructed in \cite{Kachru:2008yh}. The geometry is
\begin{equation} \label{lif}
ds^2 = -r^{2z} dt^2 + r^{2} d\vec{x}^2 + L^2 \frac{dr^2}{r^2},
\end{equation}
where $L^2$ represents the overall curvature scale, and the spacetime has $d+1$ dimensions, so there are $d_s = d-1$ spatial dimensions $\vec{x}$. The asymptotically Lifshitz solutions of a bulk gravity theory are conjectured to provide a dual holographic description of a non-relativistic field theory, with the anisotropic scaling symmetry $t \to \lambda^z t$, $x^i \to \lambda x^i$. This duality is interesting both for its potential application in condensed matter, and as an extension of our understanding of holography and the relation between field theory and spacetime descriptions. The holographic dictionary, relating bulk spacetime  quantities to field theory observables, is now fairly well-developed \cite{Kachru:2008yh,Ross:2009ar,Ross:2011gu}.  As in AdS/CFT, this identifies the leading asymptotic falloff of bulk fields with sources in the dual field theory.

An interesting early observation in AdS holography \cite{Klebanov:1999tb} was that there can be more than one conformal field theory associated to a given bulk theory, depending on the choice of boundary conditions. As first noted in \cite{Breitenlohner:1982jf} for scalar fields, for some parameter values it is possible to introduce an alternative quantisation in the bulk spacetime, where a mode which is subleading in the asymptotic expansion of the field is fixed and the leading piece is allowed to vary. This alternative quantisation leads to a second conformal field theory dual to the same spacetime, with different operator dimensions for the operators dual to the bulk fields whose boundary conditions have been changed. One can also consider mixed boundary conditions, which are dual to renormalisation group flows interpolating between the two conformal field theories, generated by a double-trace deformation of the field theory. For a scalar field, the alternative quantisation is possible when the mass of the field is in the range $m^2_{BF} < m^2 < m^2_{BF} +1$, where $m^2_{BF} = -\frac{d^2}{4}$ is the Breitenlohner-Freedman bound \cite{Breitenlohner:1982jf} for $d$ boundary dimensions.

The importance of this possibility for holography was realised in part because the dimension of the operator dual to the scalar with standard boundary conditions, $\Delta_+ = \frac{d}{2} + \sqrt{ \frac{d^2}{4} + m^2}$, was always strictly greater than the unitarity bound, $\Delta > \frac{d}{2} - 1$. With the alternative boundary condition, the scalar is dual to an operator with dimension $\Delta_- = \frac{d}{2} - \sqrt{ \frac{d^2}{4} + m^2}$, which precisely saturates the unitarity bound when $m^2 = m^2_{BF} +1$.

This analysis of alternative boundary conditions in AdS deepened our understanding of the correspondence, through an improved bulk understanding of unitarity and an understanding of the relation of double-trace and more general deformations of the field theory and boundary conditions in the bulk \cite{Witten:2001ua,Berkooz:2002ug}. The early work on scalar fields was subsequently extended to consider vector fields in \cite{Witten:2003ya,Marolf:2006nd}, and for metric perturbations in \cite{Compere:2008us}.

In the bulk spacetime, the restriction to the range $m^2 < m^2_{BF} +1$ comes from the fact that the slow fall-off mode is only normalizable with respect to the usual Klein-Gordon norm for masses in this range. Similar restrictions arise for a vector field \cite{Ishibashi:2004wx,Marolf:2006nd}. A deeper understanding of this restriction was obtained in \cite{Compere:2008us}, where it was observed that one can define an alternative norm by adding boundary terms to the Klein-Gordon current, such that the solution with Neumann boundary conditions is always normalizable. However, this norm is not positive for $m^2 > m^2_{BF} + 1$ for generic boundary metrics \cite{Compere:2008us},  and for the flat boundary metric, where the norm is positive, an IR divergence appears enforcing the unitarity bound \cite{Andrade:2011dg}.

In this paper, we aim to address the same issues in Lifshitz. We will focus primarily on investigating when alternative boundary conditions are possible from the bulk spacetime perspective. In the present paper we do this calculation for a scalar field in a fixed Lifshitz background; a companion paper to follow will consider linearised fluctuations in a theory including dynamical gravity. Our intention is to use this study to shed further light on the duality. In particular, the calculation with a scalar field will lead to a new prediction for a bound on operator dimensions in Lifshitz field theories (or at least those with gravitational duals).

After some preliminary discussion of the equation of motion in section \ref{waveeq}, we first investigate the normalizability for a scalar field on a Lifshitz background in section \ref{snorm}. We find that the slow fall-off mode is normalizable, so that alternative boundary conditions are possible, for a larger range of masses than in the AdS case:
\begin{equation} \label{mrange}
m^2_{BF} < m^2 < m^2_{BF}+ z^2,
\end{equation}
where $m^2_{BF} = -\frac{1}{4}(z + d_s)^2$ and $d_s$ is the number of spatial dimensions. The lower bound $m^2_{BF}$ is the analogue of the Breitenlohner-Freedman bound in this case. The immediate source of this wider range is that the measure in the Klein-Gordon inner product includes a factor of $r^{-z}$, so slower falloffs become normalizable as $z$ increases. More deeply, this can be related to the perspective of  \cite{Compere:2008us} by observing that as $z$ increases the dimension of boundary counterterms involving time derivatives of the boundary data increase, and the inner product needs to be modified by adding boundary terms only when the operator dimension is small enough for these terms to be relevant. In the field theory, the wider mass range \eqref{mrange} corresponds to a weaker restriction on the possible dimensions of scalar operators. Thus, the spacetime calculation predicts a reduced lower bound for the dimension of scalar operators in the field theory.

However, in section \ref{instab}, we show that the scalar field on a Lifshitz background with alternative boundary conditions can be unstable even when $m^2 > m^2_{BF}$.  In the AdS case, an instability for conformally-invariant boundary conditions is kinematically forbidden: conformal symmetry implies that the dispersion relation is $\omega = \pm k$. However for Lifshitz boundary conditions, the scaling symmetry only fixes the dispersion relation to be $\omega = \alpha k^z$ for some dimensionless parameter $\alpha$. If there is a family of modes where $\alpha$ has a positive imaginary part, they represent an instability.

We will first consider the question analytically for $z=2$, where the solution of the scalar equation can be written in terms of confluent hypergeometric functions. We show that the theory with Dirichlet boundary conditions is stable, but that the theory with Neumann boundary conditions is unstable if
\begin{equation}
m^2 > m^2_{BF} +1.
\end{equation}
We note that the instability is a UV effect, as if there is any instability then there will be exponentially growing modes for all values of $k$, and the dominant instability is associated with arbitrarily large values of $k$. Thus, we would expect this instability to appear for any asymptotically Lifshitz spacetime when we take Neumann boundary conditions for the scalar. We confirm this expectation analytically by considering the scalar field on the $z=2$ Lifshitz black hole solution introduced in \cite{Giacomini:2012hg}, where the wave equation can be solved in terms of hypergeometric functions. This makes it clear that the instability we are seeing here is not associated with the singularity in the interior of the spacetime for the pure Lifshitz solution \cite{Kachru:2008yh,Copsey:2010ya,Horowitz:2011gh}; this is a new UV pathology in these solutions. We complete our analytic discussion by giving a general argument for stability of black hole solutions with Dirichlet boundary conditions, following arguments previously given in \cite{Horowitz:1999jd,Giacomini:2012hg}.

We then extend our discussion to arbitrary values of $z$ by solving the wave equation numerically. We do numerical analysis both for the pure Lifshitz spacetime and for a spacetime with an IR cutoff, which is easier to treat numerically, using both a spectral method and shooting. We verify that the IR cutoff does not alter the instability for high momentum. We find that the theory with Neumann boundary conditions is unstable if $m^2 > m^2_{BF} +1$ for all $z$.

In the field theory dual to the alternative quantisation, the dimension is given by $\Delta_- = \frac{1}{2}(d_s+z) - \frac{1}{2} \sqrt{ (d_s+z)^2 + 4m^2}$, so $m^2 < m^2_{BF} + 1$ corresponds to
\begin{equation} \label{dbound}
\Delta \geq \frac{1}{2}(d_s+z) -1.
\end{equation}
This generalises the result $\Delta \geq \frac{d}{2} -1$ in the relativistic case (where $d=d_s+1$ is the number of spacetime dimensions). In the Lifshitz field theory,  there is no independent derivation of such a bound; in the absence of the usual state-operator map for such non-relativistic theories, there is no direct analogue of the usual argument for the unitarity bound. However, it is certainly surprising that the bound is so high; if we consider a free scalar field theory with kinetic term $(\partial_t \phi)^2$, Lifshitz scaling would require the scalar to have dimension $\Delta = \frac{1}{2}(d_s+z) -z$. This matches the bound that would be obtained from normalizability considerations alone, but the instability we find suggests that for interacting Lifshitz field theories the bound on the dimension is higher. We will discuss the field theory interpretation a little more in the conclusions in section \ref{concl}.

{\it Note added:} While this paper was in preparation \cite{Cynthia} appeared, which has considerable overlap with our analysis of normalizability in section \ref{snorm}.

\section{Scalar wave equation}
\label{waveeq}

We consider a scalar field satisfying the Klein-Gordon equation $\Box \phi - m^2 \phi =0$ in the Lifshitz spacetime \eqref{lif} with $m^2 <0$, and we neglect back-reaction on the spacetime metric\footnote{Extending the analysis to include back-reaction might be an interesting project for the future.}.  The wave equation in the Lifshitz geometry \eqref{lif} is
\begin{equation} \label{sweq}
r^{1-z-d_s} \partial_r (r^{z+d_s+1} \partial_r \phi) - (r^{-2z} \partial_t^2  - r^{-2} \partial_i^2 + m^2) \phi =0.
\end{equation}
As in AdS, the derivatives along the boundary direction have a subleading effect at large $r$, and
the asymptotic behaviour of the solutions is
\begin{equation} \label{foff}
\phi \sim \phi_+ r^{-\Delta_+} + \phi_- r^{-\Delta_-},
\end{equation}
where
\begin{equation}
\Delta_{\pm} = \frac{1}{2} (z+d_s) \pm \sqrt{ m^2 + \frac{1}{4} (z + d_s)^2}.
\end{equation}
The analogue of the Breitenlohner-Freedman bound \cite{Breitenlohner:1982jf} for Lifshitz spacetimes is thus $m^2_{BF} = -\frac{1}{4} (z+d_s)^2$. We will also write $m^2 = m^2_{BF} + \nu^2$, so $\Delta_\pm = \frac{1}{2}(z+d_s) \pm \nu$, with $\nu >0$ by convention. The usual boundary condition is to fix the slow fall-off mode $\phi_-$ and let $\phi_+$ fluctuate. We will generally refer to this as a Dirichlet boundary condition, and to the converse condition of fixing $\phi_+$ and letting $\phi_-$ fluctuate as a Neumann boundary condition, although this terminology is really only valid for $m=0$ where $\Delta_-=0$.

To solve the equation \eqref{sweq} explicitly, we use a boundary plane wave basis, writing
\begin{equation}
\phi = e^{-i \omega t + i \vec{k} \cdot \vec{x}} \psi(r).
\end{equation}
The wave equation then becomes
\begin{equation} \label{req}
r^{1-z-d_s} \partial_r (r^{z+d_s+1} \partial_r \psi) - (-r^{-2z} \omega^2 + r^{-2} k^2 + m^2) \psi =0.
\end{equation}
The $d_s$ dependence here can be simplified by writing $\psi(r) = r^{-\frac{d_s+z}{2}} \chi(r)$; then
\begin{equation} \label{ceq}
r \partial_r (r \partial_r \chi) - (-r^{-2z} \omega^2 + r^{-2} k^2 + \nu^2) \chi =0.
\end{equation}
The asymptotic behaviour of $\chi$ is then easily seen to be $\chi \sim r^{\pm \nu}$ as $r \to \infty$, with the plus (minus) sign corresponding to Neumann (Dirichlet) boundary conditions, and $\chi \sim e^{\pm i \frac{\omega}{z r^z}}$ as $r \to 0$. For real $\omega$ both behaviours at $r \to 0$ are regular. For complex $\omega$ one grows exponentially and the other decays; we select the exponentially damped mode.

We can also note that either $\omega$ or $k$ can be absorbed by a redefinition of $r$, so the equation can be rewritten as
\begin{equation}
r \partial_r (r \partial_r \chi) - \left(r^{-2} + \nu^2 - r^{-2z} \frac{\omega^2}{k^{2z}} \right) \chi =0.
\end{equation}
This makes manifest the fact that the physics on the pure Lifshitz background can depend only on the dimensionless combination $\omega/k^z$.

For generic $\omega$ and $z$, \eqref{ceq} has no solution in terms of known special functions. For $\omega =0$, we can solve it in terms of Bessel functions, but for both the normalizability and instability discussions our interest is in solutions with non-zero $\omega$. More helpfully, for $z= 2$, the equation reduces to a confluent hypergeometric equation; this was previously analysed in \cite{Kachru:2008yh}, and we will use this solution in studying the instability in section \ref{instab}.

\section{Normalizability for scalars}
\label{snorm}

In this section, we consider the normalizability of the probe scalar field. We first give a simple consideration of normalizability with respect to the normal Klein-Gordon inner product, showing that the range of masses for which alternative boundary conditions are possible is enlarged as we increase $z$. We then do a more detailed analysis of the inner product and counterterms, following \cite{Compere:2008us,Andrade:2011dg} closely. This allows us to understand the result better from the spacetime point of view, seeing the relation to kinetic counterterms in the action. We verify that inside our mass range, the standard Klein-Gordon inner product without any explicit boundary contributions is an appropriate inner product; in particular it is finite and positive definite for real $\omega$.

We assume we use the standard Klein-Gordon inner product,
\begin{equation} \label{kgn}
(\phi_1, \phi_2) = \frac{i}{2} \int_\Sigma d^{d_s} x dr \sqrt{h} n^\mu (\phi_1^* \partial_\mu \phi_2 - \phi_2 \partial_\mu \phi_1),
\end{equation}
where $\Sigma$ is a spacelike surface, which we will take to be a surface of constant $t$.
The wave equation \eqref{req} can be written as a Sturm-Liouville (SL) problem with eigenvalue $\lambda = \omega^2$ for the operator
\begin{equation}\label{SL gen}
    L = w(r)^{-1} \left[ - \frac{d}{dr} \left( p(r) \frac{d}{dr} \right ) + q(r)  \right],
\end{equation}
with $p = r^{(d_s+z+1)}$, $w = r^{d_s-z-1}$ and $q = r^{(d_s + z -1)}(m^2 + r^{-2}\vec{k}^2)$. The inner product \eqref{kgn} then becomes
\begin{equation} \label{norm}
(\phi_1, \phi_2) = (2\pi)^{d_s} \delta^{(d_s)} (\vec k_1 - \vec k_2) e^{i(\omega_1 - \omega_2) t} \frac{(\omega_1 + \omega_2)}{2} \langle \psi_1, \psi_2 \rangle_{SL},
\end{equation}
where $\langle \cdot, \cdot \rangle_{SL}$ is the corresponding SL inner product
\begin{equation}\label{SL ip}
    \langle \psi_1, \psi_2 \rangle_{SL} = \int_0^\infty r^{d_s-z-1} \psi^{\ast}_1 \psi_2 dr .
\end{equation}
With the standard Dirichlet boundary conditions $\phi_- = 0$, the fields fall off as $\phi \sim r^{-\Delta_+}$, and the large $r$ behaviour of the integral is $\int^\infty r^{-2\nu -2z -1} dr$, so the fast fall off modes are normalizable for any $\nu$. If we consider instead the Neumann boundary condition $\phi_+ = 0$, then $\phi \sim r^{-\Delta_-}$, and the large $r$ behaviour of the integral is $\int^\infty r^{2\nu -2z -1} dr$, so the slow fall off modes are normalizable with respect to this standard inner product if $\nu < z$, that is if
\begin{equation}
m^2_{BF} < m^2 < m^2_{BF} + z^2.
\end{equation}
Increasing $z$ thus increases the mass range for which the Neumann boundary conditions are allowed. In this range, we could also consider mixed boundary conditions; the flux through infinity vanishes, so that the inner product is conserved, for any linear boundary condition $\phi_+ = f \phi_-$ for real $f$.

The theory with Neumann boundary conditions is dual to a field theory with Lifshitz scaling with an operator of dimension
\begin{equation} \label{ubound}
\Delta_- = \frac{1}{2} (z + d_s) - \nu .
\end{equation}
The more general mixed boundary conditions are dual to non-scale invariant theories, which should interpolate between the Neumann boundary conditions in the UV and Dirichlet boundary conditions in the IR. We note that for $z > d_s$, $\Delta_-$ could formally take negative values. We will find an instability in the next section before we reach such values, however, so we will confine ourselves to considering $\nu < \frac{1}{2}(z+d_s)$, where a linearised analysis of the asymptotics is possible.

\subsection{Inner product and counterterms}

To gain a better understanding of the origins of the extended region of normalizability, we should consider a more careful analysis of the inner product, following \cite{Compere:2008us,Andrade:2011dg}. The argument above assumed that we could use the standard Klein-Gordon inner product. However, \cite{Compere:2008us} made it clear that the presence of kinetic boundary counterterms in the action for fields in AdS implies that we generally need to add corresponding boundary terms to the inner product to ensure that it is conserved. The correct inner product is always finite, as the boundary terms cancel any divergence from the bulk, but these terms may spoil positivity. The correct question to ask is then when we need to add boundary contributions to the Klein-Gordon inner product, and to check that the inner product is positive and conserved. We will see below that $m^2 < m^2_{BF} + z^2$ is precisely where no explicit boundary contribution is required to have a conserved inner product.

The terms $\phi_\pm$ in \eqref{foff} can be expanded in a double power series in $r^{-2}$, $r^{-2z}$,
\begin{equation}
\phi_- =  \phi^{(0)} + r^{-2} \phi^{(1)} + r^{-4} \phi^{(2)} + r^{-2z} \phi^{(z)} + r^{-2-2z} \phi^{(z+1)} + \ldots, \quad \phi_+ = \phi^{(\nu)} + \ldots,
\end{equation}
where the terms which involve powers of $r^{-2}$ are local functions of $\phi^{(0)}$ and its spatial derivatives, while the terms which involve powers of $r^{-2z}$ are functions of temporal derivatives of $\phi^{(0)}$.\footnote{We don't write the subleading terms involving derivatives of $\phi^{(\nu)}$ explicitly because they don't enter the calculation.} For example,
\begin{equation}\label{phi 1}
    \phi^{(1)} = \frac{1}{4(\nu-1)} \partial_i^2 \phi^{(0)},
\end{equation}
while
\begin{equation}\label{phi 1+eta}
    \phi^{(z)} = \frac{1}{4z (z-\nu)}\partial_t^2 \phi^{(0)}.
\end{equation}
If $\nu < z$, the terms involving temporal derivatives of $\phi^{(0)}$ are subleading compared to $\phi^{(\nu)}$ in the asymptotic expansion of $\phi$:
\begin{equation}
\phi \sim r^{-\frac{1}{2}(z+d_s)+\nu} ( \phi^{(0)} + r^{-2} \phi^{(1)} + r^{-4} \phi^{(2)} + \ldots + r^{-2 \nu} \phi^{(\nu)} + r^{-2z} \phi^{(z)} + \ldots ).
\end{equation}

To obtain a finite on-shell action, one needs to add counterterms to the bare action to cancel divergences associated to the terms $\phi^{(0)}$ and $\phi^{(i)}$ for $i < \nu$. For example, if $1 < \nu < 2$, there will be divergences in the on-shell action involving both $\phi^{(0)}$ and $\phi^{(1)}$,
\begin{eqnarray}
I_{bare} &=& - \frac{1}{2} \int_{r=r_0} d^{d_s}x dt \sqrt{-\gamma} n^\mu \phi \partial_\mu \phi \\ \nonumber & = & \frac{1}{2} \int_{r=r_0} d^{d_s}x dt r_0^{2\nu} [ \Delta_- \phi^{(0)2} + (\Delta_- + 2) \phi^{(0)} \phi^{(1)} r_0^{-2} + \Delta_+ \phi^{(0)} \phi^{(\nu)} r_0^{-2\nu} + \ldots ],
\end{eqnarray}
The divergences are cancelled by adding appropriate counterterms. The leading divergence is cancelled by a $\phi^2$ counterterm, and the $\phi^{(0)} \phi^{(1)}$ divergence can be cancelled by a $(\partial_i \phi)^2$ counterterm.

In \cite{Compere:2008us}, it was shown that derivative counterterms in the action naturally lead to boundary contributions to the symplectic structure. At the level of the inner product, we can understand these terms as being required to ensure conservation of the inner product: we add boundary terms to the Klein-Gordon current so that the flux through the boundary at large $r$ vanishes. The key difference in our case is that the counterterm in the action only involves spatial derivatives. As a result, the usual Klein-Gordon current will be conserved despite the appearance of a $\phi^{(1)}$ term in the flux through the boundary.
The flux of the Klein-Gordon current through the boundary is
\begin{eqnarray}
F &=& \frac{i}{2} \int_{r=r_0}d^{d_s}x dt \sqrt{-\gamma}  n^\mu (\phi_1^* \partial_\mu \phi_2 - \phi_2 \partial_\mu \phi_1^*) \\ \nonumber &=& \frac{i}{2} \int_{r=r_0} d^{d_s}x dt r_0^{2\nu} [2 r_0^{-2} (\phi_1^{(1)*} \phi_2^{(0)} - \phi_1^{(0)*} \phi_2^{(1)}) + (\Delta_+ - \Delta_-) r_0^{-2\nu} (\phi_1^{(\nu)*} \phi_2^{(0)} - \phi_1^{(0)*} \phi_2^{(\nu)})].
\end{eqnarray}
If we work in the plane-wave basis, the divergent term is
\begin{equation}
i \int_{r=r_0}d^{d_s}x dt r_0^{2\nu-2} \frac{k_1^2 - k_2^2} {4(1-\nu)} e^{i (\omega_1 - \omega_2) t - i (\vec{k}_1 - \vec{k}_2) \cdot \vec x} \psi_1^{(0)*}(r) \psi_2^{(0)}(r),
\end{equation}
and if we integrate over the region between two surfaces $t = $ constant, the integral over the spatial directions will introduce an overall momentum delta-function, so that this divergent term vanishes.\footnote{One can add a local counterterm to cancel this divergence; this counterterm will not affect the value of the inner product.} As claimed earlier, the finite piece will vanish for any mixed boundary condition $\phi_+ = f \phi_-$ with a real coefficient $f$. Thus, for $\nu <2$, the Klein-Gordon inner product is conserved for any such mixed boundary conditions; in particular it is conserved for both Dirichlet and Neumann boundary conditions, without adding an explicit boundary term to the inner product.

This argument can easily be extended to general $\nu < z$. The key point is that the subleading terms appearing in the asymptotic expansion for $\phi$ will all involve only spatial derivatives of $\phi^{(0)}$, so they make vanishing contributions to the total flux through the boundary at infinity between two surfaces $t = $ constant.

Once we know that the inner product is conserved, it is easy to see that it is orthogonal in the plane wave basis $\phi = e^{-i \omega t + i \vec{k} \cdot \vec{x}} \psi(r)$. We saw already in \eqref{norm} that the spatial integral makes the inner product vanish if $\vec{k_1} \neq \vec{k_2}$. Now the fact that it is independent of the spatial slice $t = $ constant we choose to evaluate it on implies it must vanish if $\omega_1 \neq \omega_2$ for real $\omega$, as otherwise it would be time dependent. We are therefore left with the inner product of the plane wave modes with themselves,
\begin{equation}
(\phi_1, \phi_1) = (2\pi)^{d_s} Vol(x) \omega_1 \int_0^\infty dr r^{d_s-z-1} |\psi_1|^2,
\end{equation}
where $Vol(x)$ is the spatial volume. This is manifestly positive if $\omega_1 >0$.

We can therefore understand the condition $\nu < z$ as arising from requiring that the free data $\phi^{(\nu)}$ appears in the asymptotic expansion before the first term which involves time derivatives of $\phi^{(0)}$. That is, it is precisely because we do not require counterterms in the action involving time derivatives that the standard Klein-Gordon norm remains appropriate for more general boundary conditions.

\section{Instability for Neumann boundary conditions}
\label{instab}

We would now like to consider the spectrum for the different possible boundary conditions, to check if there are any instabilities, looking for regular solutions which grow exponentially in time. From the point of view of the dual field theory, instabilities appear as poles in the two-point function in the upper half frequency plane.\footnote{Since our situation is time-translation invariant, poles at complex $\omega$ will appear in complex conjugate pairs.}

For the AdS case, the symmetries imply that the two-point function is a function of the Lorentz invariant $\omega^2 - k^2$. Instabilities can then occur only in the case of mixed boundary conditions, where they correspond to tachyonic poles in the two-point function with $\omega^2 -k^2 = -m^2_{bdy}$.\footnote{The mixed boundary conditions correspond to the field theory deformed by a multi-trace operator, and a bulk instability can be interpreted as this deformation making the field theory Hamiltonian unbounded from below \cite{Hertog:2004ns,Hertog:2005hm}. } For the conformally invariant pure Dirichlet or Neumann boundary conditions, by contrast, no such instability is possible, as there is no scale to provide a value for $m^2_{bdy}$.

In the Lifshitz case however, the symmetry is less restrictive, and the two-point function for general boundary conditions can depend separately on $\omega$ and $k$. For conformally invariant boundary conditions, the two-point function (up to overall scaling) must be a function of the invariant $\omega/ k^z$, but this still admits the possibility of instability, if the two-point function has a pole at
\begin{equation}
\frac{\omega}{k^z} = \alpha, \quad {\rm Im} \ \alpha > 0.
\end{equation}
We will see below that such unstable modes appear for Neumann boundary conditions when $\nu >1$. The existence of such instabilities in the conformally invariant case is perhaps surprising. Moreover, as a result of the scale invariance, the instability has no associated timescale. Unlike the relativistic case, when an instability occurs, there are exponentially growing modes for all momenta, and the modes of high momenta have arbitrarily high growth rates. Thus, this instability indicates the complete breakdown of the expansion around the putative background, which is not valid even in an open neighbourhood in time. We therefore interpret this instability as saying that no dual field theory exists for the boundary conditions where the instability is present.

Because there are unstable modes of arbitrarily high momentum, this is essentially a UV effect in the field theory. That is, this instability will affect not just the pure Lifshitz spacetime, but any solution which asymptotically approaches this solution with the Neumann boundary conditions.

In particular, this implies that the theories with mixed boundary conditions will also be unstable, as these approach the Neumann boundary condition in the UV. That is, the field theories dual to the mixed boundary conditions are relevant deformations of the scale-invariant theory dual to the Neumann boundary conditions, so the non-existence of this UV theory implies that the theories dual to the mixed boundary conditions will also not be well-defined.

\subsection{Analytic calculation for $z=2$}

We first consider the case $z=2$, where it is possible to write the solution explicitly in terms of confluent hypergeometric functions, as discussed in \cite{Kachru:2008yh}. For $z=2$, the generic solution of \eqref{req} is
\begin{equation} \label{gch}
\psi(u) = e^{iu^2 \omega/2} [\phi_+ u^{\frac{(d_s+2)}{2}+\nu} {}_1 F_1 (a,b,-iu^2 \omega)+ \phi_- u^{\frac{(d_s+2)}{2}-\nu} {}_1 F_1(a-b+1,2-b,-iu^2 \omega)],
\end{equation}
where $u= 1/r$,
\begin{equation} \label{ab}
a = \frac{1}{2}(1+\nu) + i \frac{k^2}{4 \omega}, \quad b = 1+\nu,
\end{equation}
and ${}_1 F_1 (\alpha,\gamma,z)$ is the confluent hypergeometric function, whose series expansion is ${}_1 F_1 (\alpha,\gamma,z) = 1 + \frac{\alpha}{\gamma} z + \ldots$. The first term in \eqref{gch} corresponds to the fast falloff mode at infinity, and the second term to the slow falloff.

For complex $\omega$, the solution which is regular at $r=0$ is the one which is exponentially damped. The solution which is regular for ${\rm Im \ \omega} > 0$ can be written as
\begin{equation}\label{profile tach z = 2}
    \psi = e^{i u^2 \omega/2} u^{\frac{(d_s+2)}{2}+\nu} U \left(a,b, -i \omega u^2  \right)
\end{equation}
where $u = 1/r$, the constants $a, b$ are given in \eqref{ab},  and $U(a,b,z)$ is Tricomi's confluent hypergeometric function, which is given in terms of the confluent hypergeometric function by
\begin{equation} \label{chyp}
U ( a , b, z) =  \frac{ \Gamma(1-b)}{\Gamma(a-b+1)} {}_1 F_1 (a, b, z) + \frac{\Gamma(b-1)}{\Gamma(a)}  z^{1-b} {}_1F_1(a-b+1 ,2-b, z).
\end{equation}
We then want to ask for values of $\omega$ such that this also satisfies the boundary condition at $u=0$. Using \eqref{chyp}, the solution near $u=0$ is
\begin{equation}
\psi =  u^{\frac{d_s+2}{2}-\nu} \frac{(-i \omega)^{-\nu} \Gamma(\nu)}{\Gamma\left( \frac{1+\nu}{2} + i \frac{k^2}{4 \omega} \right)} ( 1 + \ldots) + u^{\frac{d_s+2}{2}+\nu} \frac{ \Gamma(-\nu)}{\Gamma\left( \frac{1-\nu}{2}+ i \frac{k^2}{4 \omega}  \right) } (1+ \ldots),
\end{equation}
and it will satisfy Dirichlet (Neumann) boundary conditions if the first (second) term vanishes, which is when the corresponding Gamma functions have a pole. This implies that the frequencies must satisfy
\begin{equation}\label{w tach}
    \omega = - i \frac{k^2}{2(1+2n \pm \nu)} \qquad {\rm for } \ n =0,1,2,3, \ldots
\end{equation}
where the plus (minus) sign corresponds to Dirichlet (Neumann) boundary conditions.  With Dirichlet boundary conditions this condition cannot be satisfied with ${\rm Im} \ \omega > 0$ for any $n$. There are thus no solutions which satisfy Dirichlet boundary conditions and are regular on the horizon for ${\rm Im} \ \omega \neq 0$. But for Neumann boundary conditions and $\nu > 1$, there are solutions of \eqref{w tach} with ${\rm Im} \ \omega > 0$. These are therefore unstable modes which satisfy the boundary conditions both at infinity and the horizon. We see that the number of unstable modes increases as $\nu$ crosses integer values.

Note that $\omega$ is pure imaginary; unlike the usual quasi-normal modes in the relativistic case which have decaying oscillations, the unstable mode simply grows exponentially with no oscillatory component. We can thus also view this as the analytic continuation of a mode of real frequency on the Euclidean Lifshitz solution.

\subsubsection{Black hole for $z=2$}

As we argued above, this instability should appear in any asymptotically Lifshitz spacetime. One simple case where we can check this explicitly is for the black hole considered in \cite{Giacomini:2012hg}\footnote{The $d_s = 2$ version of this metric was previously found in \cite{Balasubramanian:2009rx} as a solution of Einstein gravity coupled to various matter fields. The scalar quasi-normal mode spectrum was discussed also in \cite{Balasubramanian:2009rx} for particular values of the couplings, and then generalized in \cite{Gonzalez:2012de}.}. The geometry is
\begin{equation}
ds^2 = -r^4 \left( 1- \frac{r_+^2}{r^2} \right) dt^2 + \frac{dr^2}{r^2 \left( 1- \frac{r_+^2}{r^2} \right)} + r^2 d\vec{x}^2.
\end{equation}
This metric can be obtained as a solution of a higher-derivative theory of gravity \cite{Giacomini:2012hg}, but for our present purposes the point is that it provides an example of an asymptotically Lifshitz spacetime where the wave equation can be solved analytically. For this metric the wave equation in the boundary plane wave basis becomes
\begin{equation}
r^{-d_s -1} \partial_r (r^{d_s +1} (r^2 -r_+^2) \partial_r \psi) - \left( m^2 + \frac{k^2}{r^2} + \frac{\omega^2}{r^2 (r^2-r_+^2)} \right) \psi = 0.
\end{equation}
Introducing a radial variable $y = \frac{r^2 - r_+^2}{r^2}$, this becomes a hypergeometric equation \cite{Balasubramanian:2009rx}. The solution which is well-behaved at the black hole horizon $y =0$ for ${\rm Im}\  \omega >0$ is
\begin{equation} \label{bhsoln}
\psi = y^{-i \frac{\omega}{2 r_+^2}} (1-y)^{\frac{1}{4}(d_s +2) + \frac{1}{2} \nu} F(\alpha,\beta,\gamma; y),
\end{equation}
where $F(\alpha,\beta,\gamma; y)$ is the hypergeometric function and
\begin{equation}
\alpha, \beta = \frac{1}{2} \left( 1 + \nu - \frac{i\omega}{r_+^2} \pm \sqrt{ \frac{d_s^2}{4} - \frac{\omega^2}{r_+^4} - \frac{k^2}{r_+^2} } \right), \quad \gamma = 1 - \frac{i\omega}{r_+^2}.
\end{equation}
Analytically continuing the hypergeometric function to $y=1$ gives
\begin{eqnarray} \label{hexp}
F(\alpha,\beta,\gamma; y) &=& \frac{\Gamma(\gamma) \Gamma(\gamma-\alpha-\beta)}{\Gamma(\gamma-\alpha) \Gamma(\gamma-\beta)} F(\alpha,\beta,\alpha+\beta-\gamma+1, 1-y) \\ && \nonumber + (1-y)^{\gamma -\alpha -\beta} \frac{\Gamma(\gamma) \Gamma(\alpha+\beta-\gamma)}{\Gamma(\alpha) \Gamma(\beta)} F(\gamma-\alpha,\gamma-\beta,\gamma-\alpha-\beta+1, 1-y).
\end{eqnarray}
Thus, the solution will satisfy Dirichlet boundary conditions at infinity if the second term vanishes, which is to say if there are poles in the Gamma functions in the denominator, $\alpha = -n$ or $\beta = -n$ for some integer $n$. Rearranging and squaring gives
\begin{equation}
2i \omega (1 + \nu + 2n)  = k^2 + r_+^2 \left( (1+\nu + 2n)^2 - \frac{d_s^2}{4} \right).
\end{equation}
We see that for large $k$, the term proportional to $r_+^2$ becomes negligible, and the spectrum reduces to the one found in \eqref{w tach}, as expected. There is thus no instability at large $k$; however it is interesting to note that for $d_s >2$, the black hole can produce instabilities for {\it small} $k$ if $\nu$ is small enough. The black hole will have at least one exponentially growing mode if $(1+\nu)^2 - \frac{d_s^2}{4} < 0$.\footnote{This instability was missed in \cite{Giacomini:2012hg} because they only considered Dirichlet boundary conditions with $m^2 \geq 0$, corresponding to $\nu \geq \frac{1}{2} (d_s +2)$.}

The solution \eqref{bhsoln} will satisfy Neumann boundary conditions if the first term in \eqref{hexp} vanishes, that is if $\gamma - \alpha = -n$ or $\gamma -\beta =-n$ for some integer $n$. As in the pure Lifshitz spacetime this simply corresponds to $\nu \to - \nu$ in the Dirichlet analysis above. The spectrum of such modes is given by
\begin{equation}
2i \omega (1 - \nu + 2n)  = k^2 + r_+^2 \left( (1- \nu + 2n)^2 - \frac{d_s^2}{4} \right).
\end{equation}
Again, this reduces to \eqref{w tach} at large $k$. We see explicitly that the unstable modes at large $k$ are present for the black hole solution as well.

\subsection{Stability for Dirichlet boundary conditions for general $z$}

For general $z$, we cannot solve the wave equation analytically. However, for Dirichlet boundary conditions, we can give a general argument that no such instability can occur. This argument does not consider the pure Lifshitz spacetime directly; instead it is convenient to regulate the IR behaviour by considering an asymptotically Lifshitz black hole solution and using an argument for stability originally developed for AdS black holes in \cite{Horowitz:1999jd} and extended to the Lifshitz case in \cite{Giacomini:2012hg}. The absence of an instability for large $k$ on the black hole spacetime will then imply the absence of any instability on the pure Lifshitz spacetime.

We consider the black hole in ingoing Eddington-Finkelstein coordinates, where we assume the metric is of the form
\begin{equation}
ds^2 = -r^{2z} f(r) dv^2 + 2 r^{z-1} dv dr + r^2 d\vec{x}^2,
\end{equation}
where $f(r)$ is some emblackening function with $f(r_+) = 0$, $f'(r) > 0$ for $r \geq r_+$, and $f(r) \to 1$ as $r \to \infty$.  As we are introducing the black hole simply as a convenient IR regulator, we do not specify a precise form for $f(r)$ or consider the equations of motion for the metric. We will take the scalar field to be of the form $\phi = e^{-i\omega v + i \vec k \cdot \vec x} r^{-\frac{d_s}{2}} R(r)$. Then the wave equation becomes
\begin{equation} \label{inwe}
(r^{z+1} f(r) R')' - 2i \omega R' - V(r) R = 0,
\end{equation}
where
\begin{equation}
V(r) = k^2 r^{z-3} + \left( \nu^2 - \frac{z^2}{4} - \frac{d_s(2z+d_s)}{4} (1- f(r)) + \frac{d_s}{2} r f' \right)r^{z-1}.
\end{equation}

To argue for stability, we want to see that ${\rm Im} \ \omega \leq 0$. By multiplying \eqref{inwe} by $R^*$ and integrating, we can see that
\begin{equation}
\int_{r_+}^\infty dr [ R^* (r^{z+1} f(r) R')' - V(r) |R|^2 - 2i \omega R^* R'] = 0.
\end{equation}
Thus, integrating by parts gives that
\begin{equation}
X = \int_{r_+}^\infty dr [ r^{z+1} f(r) |R'|^2 + V(r) |R|^2] = - 2i \omega \int_{r_+}^\infty dr  R^* R'  + r^{z+1} f(r) R^* R' |_{r_+}^\infty.
\end{equation}
Now $f(r_+) = 0$ ensures that the boundary term at the horizon vanishes (this is why it is useful to introduce a black hole in the bulk), and at $r \to \infty$, the solution of \eqref{inwe} asymptotically behaves as $R \sim r^{-\frac{z}{2} \pm \nu}$. The lower sign is the fast fall-off mode which is retained for Dirichlet boundary conditions, so for Dirichlet boundary conditions, the boundary term at infinity also vanishes. Thus,
\begin{equation}
X = - 2i \omega \int_{r_+}^\infty dr  R^* R' .
\end{equation}
Since $X$ is real, taking the imaginary part of this equation gives $(\omega - \bar \omega)  \int_{r_+}^\infty dr R^* R' = \bar \omega |R|^2_{r=r_+}$, so
\begin{equation}
X = - \frac{|\omega|^2}{{\rm Im} \ \omega} |R|^2_{r=r_+}.
\end{equation}
So if $X \geq 0$, ${\rm Im} \ \omega \leq 0$ and the mode is stable.

For sufficiently large $\nu$,  $V(r)$ is positive everywhere, and $X$ is obviously positive. However, we wish to give an argument for all positive $\nu$. This is possible because we are concerned only with instabilities which continue to appear for arbitrarily large $k$: only such instabilities can correspond to instabilities of the pure Lifshitz spacetime. We can therefore focus on the region at large $r$ in the integral defining $X$; if there is a problem at large $k$ it must come from this part where the $k^2$ contribution in $V$ is suppressed relative to the other terms. The large $r$ part of $X$ is approximately
\begin{eqnarray}
X &\sim& \int^\infty dr [ r^{z+1} |R'|^2 + r^{z-1} (\nu^2 - \frac{z^2}{4}) |R|^2] \nonumber \\
  &\sim& \int^\infty dr [(\nu + \frac{z}{2})^2 + (\nu^2 - \frac{z^2}{4})] r^{-2 \nu -1} \nonumber \\
  &\sim& \int^\infty dr [2 \nu^2 + \nu z] r^{-2 \nu -1},
\end{eqnarray}
where we used the falloff $R \sim r^{-\frac{z}{2} - \nu}$ for Dirichlet boundary conditions again in the second line. Thus this large $r$ contribution is positive and bounded, and at least for sufficiently large $k$, $X \geq 0$ and there will be no instability. We do not exclude the possibility of an instability at finite $k$ -we saw in the previous section that such instabilities can in fact occur - but the absence of instability at arbitrarily large $k$ implies the pure Lifshitz solution with Dirichlet boundary conditions cannot be unstable.

\subsection{Numerical analysis for Neumann boundary conditions for general $z$}

Having seen that the theory with Dirichlet boundary conditions is stable, we now turn to a numerical analysis to determine when the theory with Neumann boundary conditions is unstable. The numerical determination of the spectrum is conceptually the same as the analytic calculation for $z=2$: we choose a solution which is regular in the interior and look for values of $\omega$ which satisfy the asymptotic Neumann boundary conditions. This defines an eigenvalue problem for $\omega$, which we can solve either by spectral methods or by shooting.

For integer $z$, we can apply spectral methods to solve the problem in the pure Lifshitz geometry, as the solution is analytic even near $r=0$ after appropriate redefinitions. For non-integer values of $z$, we need to modify the problem, as the numerical analysis for pure Lifshitz is difficult because of the behaviour near the horizon. Fortunately, we have seen that the instability that we are interested in is essentially a UV effect in the field theory, so we can modify the geometry in the interior of the spacetime without affecting the instability. We could consider a black hole solution, as we did previously, but for numerical studies it is more convenient to simply cut off the spacetime at a hard wall. We introduce a radial cutoff, rescaling the radial coordinate so that the cutoff is at $r = 1$ and consider the region $r > 1$. For simplicity, we focus on Dirichlet or Neumann boundary conditions at the wall; these two choices lead to similar results, and give the same region of instability in $\nu$.

\begin{figure}[h]
\begin{center}
\includegraphics[scale=0.75]{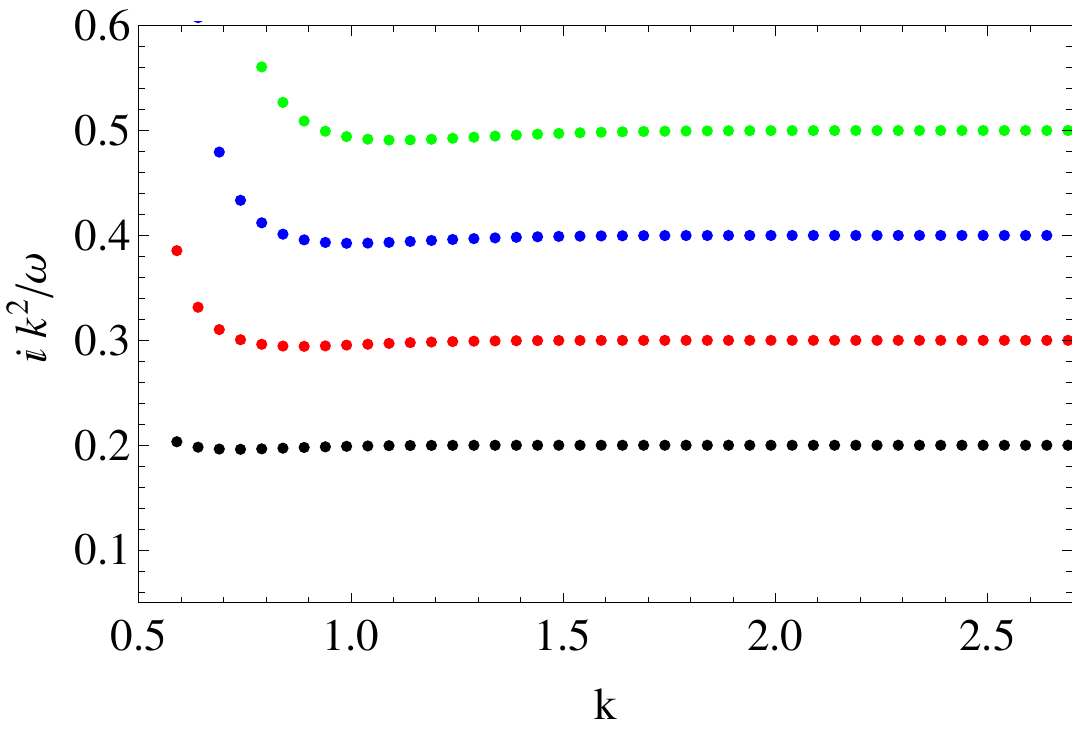}
\caption{Unstable modes for $z=2$ for Neumann boundary conditions at the IR wall, as a function of $k$ for various values of $\nu$. The lines of data correspond to $\nu = 1.1$, $\nu = 1.15$, $\nu = 1.20$ and $\nu = 1.25$, with $\nu$ increasing as we move up the plot. We observe that for sufficiently large $k$, the relation \eqref{w tach} is reproduced. }
\label{IR wall z=2}
\end{center}
\end{figure}

We have done the numerical analysis for the theory with an IR wall using both spectral methods and shooting.  As the wall breaks the radial scaling symmetry, we have to consider varying $\omega$ and $k$ separately. We are interested in solutions at large $k$, giving instabilities of the pure Lifshitz spacetime. We solve the eigenvalue problem for $\omega^2$ for fixed $k$, and check that $k^z/\omega$ is invariant as we vary $k$, for relatively large $k$. Failure to satisfy this property is interpreted as a sign that the numerics are inaccurate.

\begin{figure}[h]
\begin{center}
\includegraphics[scale=0.75]{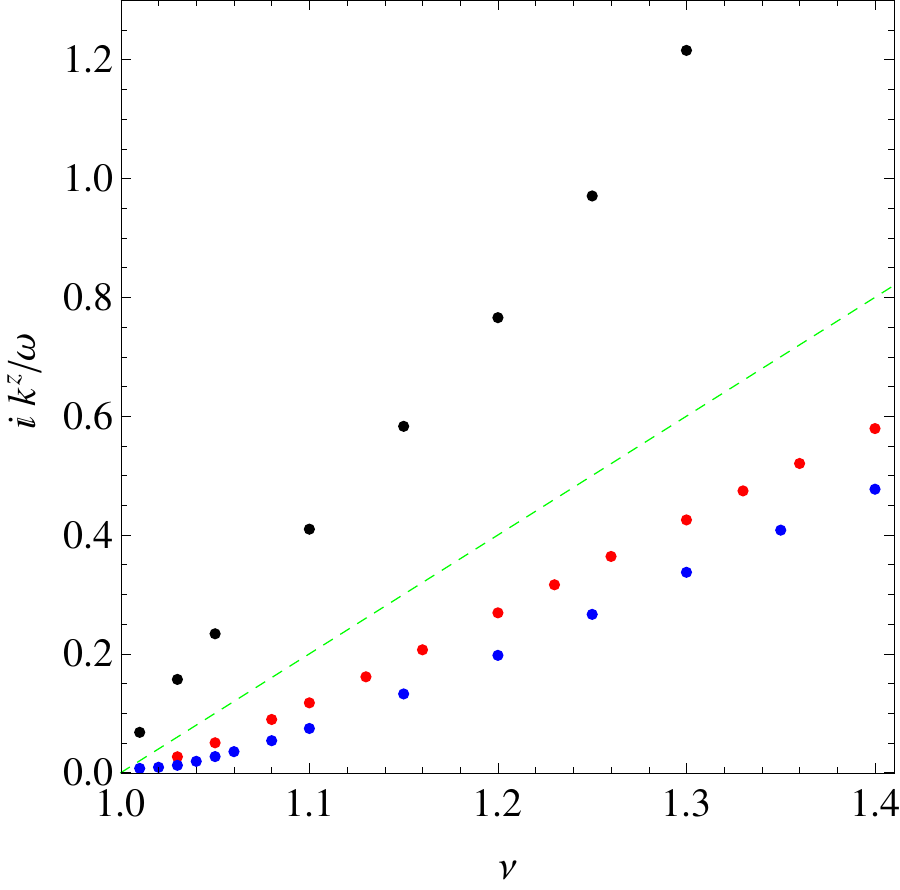}
\caption{Scale invariant combination $i k^z /\omega$ for the first unstable mode as a function of $\nu$. The top (black) data is $z = 3/2$, the dotted curve (green) is the analytic result for $z = 2$, the lower data are $z=5/2$ (red) and the lowest line of data (blue) is $z=3$. The numerics become inaccurate for large $\nu$, problems appearing roughly for $\nu > 1.4$ for $z >2$. These are identified by varying the size of the grid and noticing that the eigenvalues change considerably.}
\label{IR wall gen z}
\end{center}
\end{figure}

\begin{figure}[h]
\begin{center}
\includegraphics[scale=0.75]{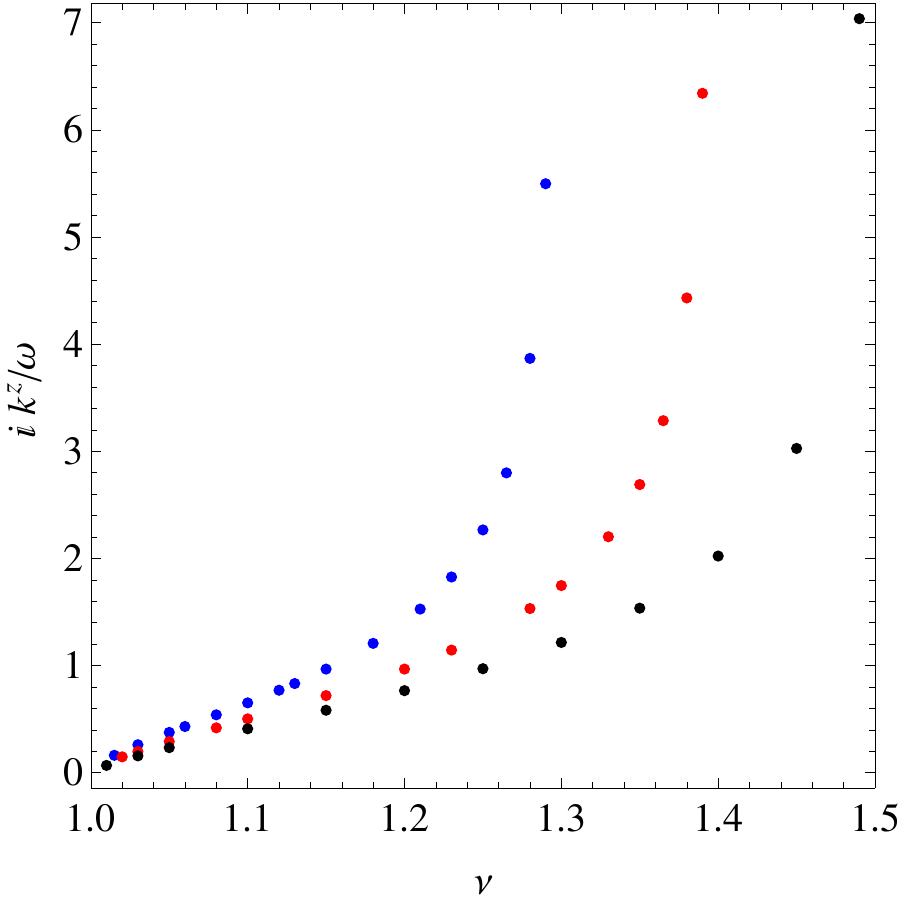}
\caption{We plot $i k^{z}/\omega$ for large $k$ as a function of $\nu$ for $z= 1.3$ (top), $z = 1.4$ (center) and $z = 1 .5$ (bottom), using shooting with an IR wall. Note that $i k^{z}/\omega$ diverges as $\nu$ approaches $z$ from below.}
\label{shooting}
\end{center}
\end{figure}

As an illustration, the values obtained by shooting for $z=2$ are plotted in figure \ref{IR wall z=2}.\footnote{Note that even though the radial equation can be solved analytically in this case, the spectrum for the hard wall must be determined numerically since the boundary condition at the wall involves (roughly speaking) determining zeros of the confluent hypergeometric function.} We see that $k^2/\omega$ does indeed become constant at large $k$, and that for large values of $k$, the relation \eqref{w tach} is accurately reproduced. In particular, as we decrease $\nu$ towards $\nu = 1$, the approximately invariant quantity $k^2/\omega$ goes to zero. Representative plots for different values of $z$  using spectral methods are given in figure \ref{IR wall gen z}. Plots obtained by shooting are given in figure \ref{shooting}. We see clearly that there is an instability for $\nu >1$.

A significant difference between the numerical results and the behaviour of the analytic solution for $z=2$ is the behaviour when $\nu \geq z$ for $z < 2$. We find that as $\nu \to z$, the invariant $\omega/ k^z$ calculated for large $k$ goes to zero. This can clearly be seen in the plots generated by shooting in figure \ref{shooting}: $k^z/\omega$ diverges as $\nu \to z$ (similar results are obtained by spectral methods for $z = 1.5$).  The general behaviour we find is that there are no complex frequency modes at large $k$ for $\nu > z$, the only mode solutions are those with real frequencies. There are however complex frequency unstable modes at small $k$ in the IR wall geometry.

In the limit as $z \to 1$, all the frequencies smoothly approach those in the IR wall cutoff version of AdS; in particular, there is a tachyonic instability in the IR cutoff AdS with Neumann boundary conditions whose scale is set by the location of the wall, and this is reproduced by the behaviour of the complex frequency unstable solutions at small $k$ for $\nu > z$. Some representative large $k$ values are plotted in figure 4.

\begin{figure}[htb]
\center
\subfigure[][]{
\label{w vs z 1}
\includegraphics[width=0.4\linewidth]{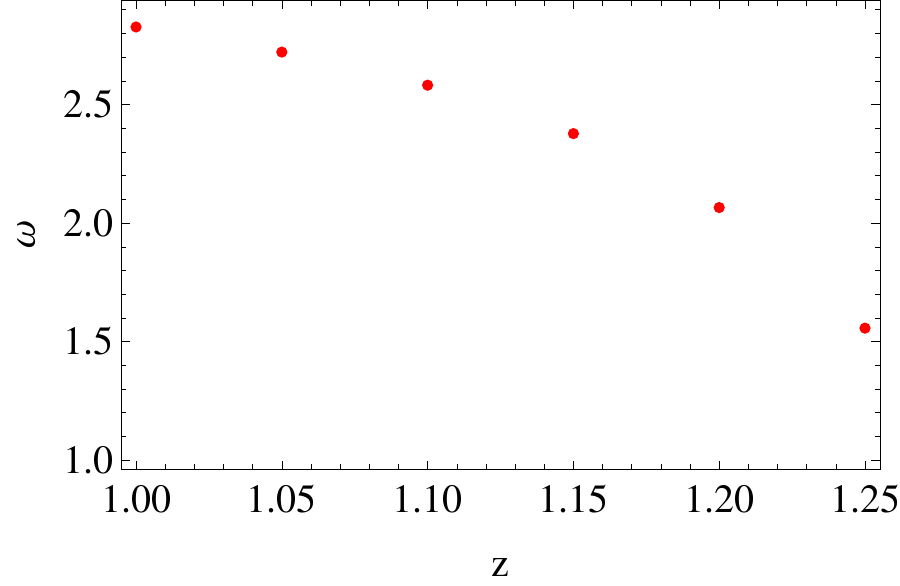}}
\qquad \qquad
\subfigure[][]{
\label{w vs z 2}
\includegraphics[width=0.4\linewidth]{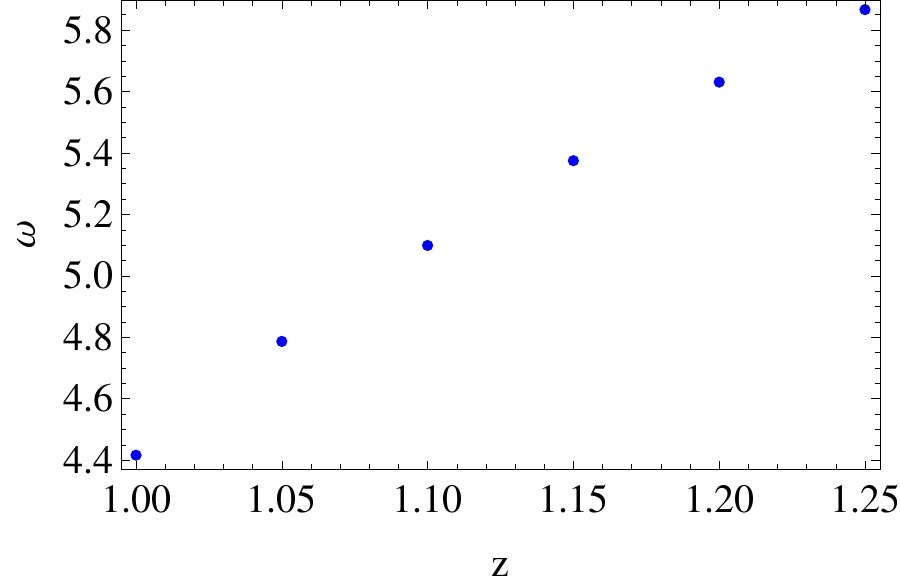}}
\caption{We plot the frequencies of the first two modes for $\nu = 1.3$, $k = 3.0$ as a function of $z$ with $z < \nu$. All the frequencies are real and they approach the relativistic values as $z \rightarrow 1$.}
\end{figure}

This explains the relation to the relativistic case: if we were to consider varying $z$ at fixed $\nu$, the UV instability will disappear at $\nu = z$, and in the region near $z=1$ the qualitative behaviour is similar to the AdS case.  Note that while the UV instability disappears, since $\nu > z$, we are violating the normalizability bound and there will be pathologies for Neumann boundary conditions along the lines of the analysis in \cite{Andrade:2011dg}. In particular, there will be ghosts in the spectrum in the IR wall geometry.

\section{Discussion}
\label{concl}

In this paper, we have considered the possibility of alternative boundary conditions for scalar fields in the Lifshitz spacetime. We found that the usual normalizability condition implied that Neumann boundary conditions are possible for a larger range of masses than in the AdS case. However, we found that there is a new instability for scalar fields with Neumann boundary conditions on Lifshitz, which implies that there is a well-defined dual field theory only for $m^2 < m^2_{BF} +1$. This instability is a UV effect; we saw explicitly that it is not tied to the presence of a singularity at $r \to \infty$ in the Lifshitz geometry as the same instability appears in an asymptotically Lifshitz black hole and in a spacetime with an IR hard wall cutoff.

This led to the proposed lower bound \eqref{ubound} on the dimension of scalar operators in strongly-coupled field theories with Lifshitz scaling symmetry. It would be very interesting to understand this bound better from the dual field theory perspective. The bound is non-trivial: the field theory of a free scalar field, with the kinetic term $ \int dt d^{d_s} x  (\partial_t \phi)^2 $, will be invariant under a Lifshitz scaling with dynamical exponent $z$ if the scalar field has dimension $\Delta_\phi = \frac{1}{2}(d_s -z)$. This free scalar value violates our bound  \eqref{ubound},\footnote{Although intriguingly it precisely saturates the bound at which normalizability fails to apply, when $m^2 = m^2_{BF} + z^2$ in the bulk.} so if the proposed bound is to be valid, the interacting nature of the theory must play a central role.

The bound for $z=1$ corresponds to the unitarity bound in the field theory. Could there be a similarly general derivation for $z > 1$? The usual argument for the unitarity bound is based on considering the radial Hamiltonian and relating the dimensions of operators to the norms of states. With Lifshitz symmetry there are no special conformal transformations, so we cannot construct such a state-operator mapping. However, a different derivation of the unitarity bound which uses the optical theorem, and does not rely directly on a state-operator mapping, was given in \cite{Grinstein:2008qk}. This argument could in principle be extended to the Lifshitz case to obtain a bound directly from the field theory point of view, as was recently done for time-dependent cases in \cite{Dong:2012en}.  Unfortunately, to do this in practice seems technically challenging. The argument was based on relating the imaginary part of the two-point function of an operator $\mathcal O$ to the optical theorem for some external field coupled to $\mathcal O$. It is possible to calculate the two-point function in momentum space for the case $z=2$. But to apply the argument of  \cite{Grinstein:2008qk}, we need to control the overall normalisation of this two-point function, as function of $m^2$ and $z$. In \cite{Grinstein:2008qk}, the overall normalisation is fixed by requiring that $\langle \mathcal O^\dagger (x) \mathcal O(y) \rangle$ is positive definite at spacelike separation. Unfortunately, it is not possible to analytically compute the integral required to evaluate the two-point function obtained for $z=2$ in position space. Further exploration of the field theory is left as an open problem for the future.

It would also be interesting to explore these questions of alternative boundary conditions and instability for the full theory with dynamical gravity. We intend to explore these issues for linearised fluctuations of the metric and matter fields in the context of the massive vector model of \cite{Taylor:2008tg} in the future.

\section*{Acknowledgements}

We are grateful for useful conversations with Shamit Kachru, Don Marolf, Mukund Rangamani, Jorge Santos, Eva Silverstein, David Tong and Benson Way. This work is supported in part by the STFC, by the National Science Foundation under Grant No. NSF PHY11-25915 and Grant No PHY08-55415,
and by funds from the University of California. T.A. is also pleased to thank DAMTP, Cambridge, for their hospitality during the completion of this work.

\bibliographystyle{utphys}
\bibliography{lifshitz}

\end{document}